\documentclass{ws-mpla}
\newcommand{\nn}{\nonumber}

\newcommand{\be}{\begin{equation}}
\newcommand{\ee}{\end{equation}}
\newcommand{\bea}{\begin{eqnarray}}
\newcommand{\eea}{\end{eqnarray}}

\newcommand{\numu}{\mbox{$\nu_{\mu}$}}
\newcommand{\numubar}{\mbox{$\overline{\nu}_{\mu}$}}
\newcommand{\nue}{\mbox{$\nu_{e}$}}
\newcommand{\nuebar}{\mbox{$\overline{\nu}_{e}$}}

\newcommand{\tetaot}{\mbox{$\theta_{13}$}}

\newcommand{\tetatt}{\mbox{$\theta_{23}$}}

\newcommand{\delot}{\mbox{$\Delta_{23}$}}

\newcommand{\sol}{\mbox{${\Delta m^2_{12} L \over 4 E}$}}
\newcommand{\atmos}{\mbox{${\Delta m^2_{23} L \over 4 E}$}}

\begin{document}

\markboth{Unveiling neutrino mixing}
{Unveiling neutrino mixing}

%
%

\title{UNVEILING NEUTRINO MIXING AND LEPTONIC CP VIOLATION}

\author{\footnotesize OLGA MENA}

\address{Fermilab National Accelerator Laboratory\\
P.O. Box 500, Batavia, IL 60510 US\\
omena@fnal.gov}

\maketitle

\pub{Received (Day Month Year)}{Revised (Day Month Year)}

\begin{abstract}
    We review the present understanding of neutrino masses and mixings, discussing what are the unknowns in the three family oscillation scenario. Despite the anticipated success coming from the planned long baseline neutrino experiments in
 unraveling the leptonic mixing sector, there are two important unknowns which may remain obscure: the mixing angle $\theta_{13}$ and the CP-phase $\delta$. The measurement of these two parameters has led us to consider the combination of superbeams and neutrino factories as the key to unveil the neutrino oscillation picture.

\end{abstract}

\ccode{PACS Nos.: 14.60Pq}

\section{Introduction: status of neutrino masses and mixing, and pending questions.}	

\begin{center}
\begin{figure}[t]
\centerline{\psfig{file=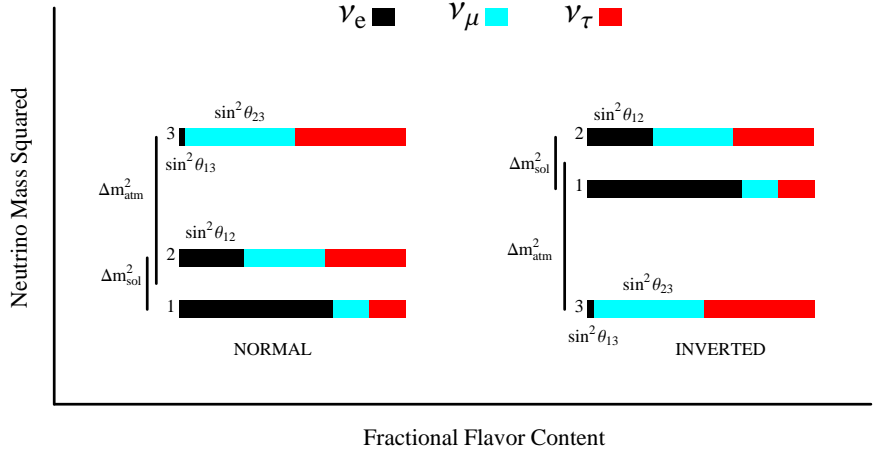, width=12cm}}
\vspace*{8pt}
\caption{\it The range of probability of finding the $\alpha$-flavor in the i-the mass eigenstate as indicated for the two different mass hierarchies for the present best 
fit values of  the mixing parameters.}
\label{fig:hierarchy2}
\end{figure}
\end{center}

Neutrino oscillations have been observed and robustly established by the data from solar~\cite{solar,sks}, atmospheric~\cite{SK}, reactor~\cite{kam} and long-baseline neutrino experiments~\cite{k2k}. Barring exotic interactions, these results indicate the existence of non-zero neutrino masses and mixings. The most economical, trivial way to accommodate these new parameters is
 via the three neutrino PMNS mixing
 matrix~\footnote{We restrict ourselves to a three-family neutrino scenario analysis. The unconfirmed LSND signal cannot be explained in terms of
 neutrino oscillations within this scenario, but might require
 additional light sterile neutrinos or more exotic explanations
 ~\cite{gabriela}.
 The ongoing  MiniBooNE experiment~\cite{miniboone} is expected to explore all
 of the LSND oscillation parameter space~\cite{LSND}.},
 the leptonic analogue to the CKM matrix in the quark sector. Neutrino oscillations within this scenario are described by six parameters: two mass squared differences\footnote{$\Delta m_{ij}^{2} \equiv m_j^2 -m_i^2$ throughout the paper.} ($\Delta m_{12}^2$ and $\Delta m_{23}^2$), three Euler angles ($\theta_{12}$, $ \theta_{23}$ and $\theta_{13}$) and one Dirac CP phase $\delta$. 
The standard way to connect the solar, atmospheric, reactor and accelerator data with some of the six oscillation parameters listed above is to identify the two mass splittings and the two mixing angles which drive the solar and atmospheric transitions with ($\Delta m_{12}^2$, $\theta_{12}$) and ($|\Delta m_{23}^2|$, $\theta_{23}$), respectively. The sign of the splitting of the atmospheric state $\Delta m_{23}^2$ with respect to the solar doublet is one of the unknowns within the neutrino sector. Consequently, the mass eigenstates $\nu_1$ and $\nu_2$ involved in the solar doublet could have smaller mass than the third mass eigenstate $\nu_3$ (normal hierarchy) or larger mass than the former doublet (inverted hierarchy). Both possibilities are illustrated in Fig.~\ref{fig:hierarchy2}, extracted from Ref~\cite{pom1}.
The best fit point for the combined analysis of solar neutrino data together with  KamLAND reactor data~\cite{kamland} is at $\Delta m_{12}^2=8.2 \times 10^{-5}$ eV$^2$ and $\tan^{2}\theta_{12}=0.4$. In the atmospheric neutrino sector, the most recent analysis of K2K accelerator neutrino data and atmospheric neutrino data~\cite{paris} finds the best fit at $|\Delta m_{23}^2|=2.7 \times 10^{-3}$ eV$^2$ and $\sin^{2}2\theta_{23}=1$.

The mixing angle $\theta_{13}$ (which connects the solar and atmospheric neutrino realms) and the amount CP violation in the leptonic sector are undetermined.
At present, the upper bound on the angle $\theta_{13}$ coming from CHOOZ reactor neutrino data~\cite{chooz} is $\sin^{2} 2 \theta_{13} < 0.1$ (at $90\%$ CL) for a value of the atmospheric mass gap close to the best fit reported before.  
The CP-phase $\delta$ is unobservable in current neutrino oscillation experiments. The experimental discovery of the existence of CP violation in the leptonic sector, together with the discovery of the Majorana neutrino character would point to leptogenesis as the source for the baryon asymmetry of the universe, provided that accidental cancellations are not present. 

Several oscillation experiments that exploit neutrino beams from nuclear reactors and accelerators are taking data, and similar experiments will take data over the next few years. All of them have inaugurated a precision era in neutrino physics. Despite the anticipated progress coming from short-term future facilities, there are fundamental issues that will remain unknown, namely, in the oscillation picture:
\begin{itemlist}
\item The value of $\theta_{13}$. Apart from the measurements of $\theta_{13}$ coming from future long baseline experiments, future reactor neutrino oscillation experiments could set the value of $\theta_{13}$, as explored in detail through Ref~\cite{white}. Also, if neutrinos from astrophysical sources decay~\cite{bellbeacom2}, the deviations of the flavor ratios with respect to the values expected in the standard oscillation scenario could allow a measurement of $\theta_{13}$ and $\delta$. 
\item The ordering of the mass spectrum.  
\item The existence of leptonic CP Violation.
\end{itemlist}

Moreover, in spite of the bright expectations associated even with the
 longer-term neutrino oscillation experiments, two fundamental neutrino properties would still remain unknown: their nature (Dirac or Majorana) and the absolute value of their masses. Both are outstanding questions that would remain to be answered with experiments other than the ones measuring neutrino
 oscillations. 

Direct information on the absolute scale of neutrino masses can be extracted from the kinematics of weak decays involving neutrinos in their final states. The present upper bound on the electron neutrino mass from tritium beta decay experiments is $2.2$ eV ($95 \%$ CL)~\cite{tritium}. The authors of Ref~\cite{tropa} find an upper limit of $1.8$ eV at the $95\%$ CL after the combination of the data from Mainz and Troitsk experiments. 
In the future, the KATRIN experiment is expected to be sensitive to electron neutrino masses of $\sim0.2$ eV at the $90\%$ CL~\cite{katrin}. 
 
Neutrinoless double beta decay searches are, at present, the only experimental way to establish the neutrino character. This type of decay is only
 possible if neutrinos are Majorana particles.
 Experimentally, the lack of the observation of neutrinoless double beta decay processes so far provides an upper bound on the so-called \emph{``effective Majorana mass''} of the electron neutrino within the range $0.3-1.0$ eV~\cite{pdg}.

Next generation of neutrinoless double beta decay experiments could provide a definitive answer to the question of whether neutrinos are Dirac or Majorana. For a complete discussion of the potential of these future experiments in measuring
 fundamental neutrino properties, see Ref~\cite{carlos}. 

Cosmology provides the key to tackle the ``bare'' absolute scale of neutrino masses, since it is sensitive to the overall neutrino mass scale, free of mixing parameters or CP-violating phase dependencies. 

As pointed out in Ref~\cite{sergio1}, neutrinos with masses smaller than the temperature at the recombination era ($T\sim 0.3$ eV) are still relativistic at the time of last scattering, and their effect in terms of CMB perturbations 
is equivalent to the one present in the massless neutrino situation. If the neutrino mass is close to $0.3$ eV, the free streaming scale is imprinted in the perturbations which enter inside the Hubble radius before decoupling. Perturbations smaller than the free streaming scale are therefore suppressed, leading to a suppression of of the matter power spectrum proportional to the ratio of the neutrino energy density to the overall matter density.

While CMB data alone does not provide a competitive bound on neutrino mass~\footnote{The free streaming scale for a neutrino mass $\sim$ eV is smaller than the ones which can be probed using present CMB data.} if compared with present terrestrial kinematic constraints,
 a stringent bound on the sum of neutrino masses of about
 $\sum_i m_{\nu_i} \le 1$ eV can be obtained
 by combining CMB data with large scale structure from 2df or SDSS~\cite{hannestad}. The upper limit on $\sum_i m_{\nu_i}$ that can be achieved with
 cosmology depends somewhat on the priors and the data exploited in
 the analysis. 
 For example, a recent joint analysis of CMB + SN-Ia + HST + LSS data~\cite{tropa} provides a $2 \sigma$ bound on $\sum_i m_{\nu_i} \le 1.4$ eV~\footnote{If Ly$\alpha$ Forest is added as additional information in the SDSS data, the previous bound is improved by a factor of $\sim 3$. However, the systematics of the Ly$\alpha$ analysis is unclear and it has to be further explored.}.
\newline
\vspace{0.5cm}

The determination of the fermion masses and mixings parameters is a mandatory first step that is essential for an understanding of the origin of flavor. Furthermore, as mentioned before, neutrino masses point to leptogenesis as the source of the matter-antimatter asymmetry of the universe, provided CP is violated in the leptonic sector.
-
 In the present study  we concentrate on the determination of neutrino oscillation parameters and on the measurement of leptonic CP violation with
 long-baseline, accelerator-based neutrino oscillation experiments. 

In particular, in the next sections we will explore in detail the
 possible measurement of the two unknown parameters $\theta_{13}$ and $\delta$ with future {\it superbeam} (SB) and {\it neutrino factory} (NF) facilities,
 as these two designs appear to be among the most promising ways
 to unveil neutrino mixing and leptonic CP violation \footnote{The prospects
 of future measurements of these two oscillation parameters at different $\beta$ beams setups has been recently explored in Ref~\cite{pilar}.}.

What is a Superbeam (SB) experiment~\cite{sb,mezzeto,jj}? It consists, basically, of a higher intensity version of a conventional neutrino (antineutrino) beam. Superbeams represent the logical next step in accelerator-based neutrino physics. A conventional neutrino beam is produced by a primary proton beam which hits a target and creates secondary beams of charged pions and kaons. If positively charged pions and kaons have been focused and
 directed into the decay channel, the resulting beam will contain mostly muon neutrinos produced in the two body decays $\pi^{+}\to\mu^{+}\numu$ and $K^{+}\to\mu^{+}\numu$. In the present work we exploit the experimental setup presented in Ref~\cite{jj}, i.e, the CERN SPL superbeam with a $400$ Kton Water Cherenkov detector~\footnote{It may be argued that such a detector may be unrealistic in practice. We shall use it with the purpose of illustrating the far-future physics perspective.} located at a $130$ km distance from the neutrino source, in the Fr\'ejus tunnel. We have assumed a $2$ year-long run with $\pi^{+}$ focusing, and a $8$
 year-long run with $\pi^{-}$ focusing. The fluxes and detector systematics have been discussed in Ref~\cite{jj}. We have not exploited the energy dependence of the signal because the neutrino spectrum is peaked at $\sim 0.25$ GeV, and  at such low energies the neutrino energy can not be reconstructed due to Fermi motion of nucleons inside the nuclear target (oxygen, in this case). 

What is a Neutrino Factory (NF)~\cite{history,geer,dgh,yrep,yrep2}? It consists, essentially, of a muon storage ring\footnote{This muon storage ring is an essential stepping-stone towards
  possible muon colliders.} with long straight sections along which the muons decay. These muons  provide high intensity neutrino beams (the neutrino flux is approximately $10^4$ times the flux of
existing neutrino beams), which have a precisely-known neutrino flavor content, making them extremely superior to the conventional beams.
Hence, compared to conventional neutrino beams from pion decay, the neutrino factory provides $\nue$ and $\nuebar$ beams in addition to $\numu$ and $\numubar$ beams, with minimal systematic uncertainties on the neutrino flux and spectrum~\cite{alicia}. 
What is a ``wrong sign muon'' event~\cite{golden}? Suppose, for example, that positive charged muons have been stored in the ring. These muons will decay as $\mu^{+}\to e^{+} + \nue + \numubar$. The muon antineutrinos will interact in the detector to produce positive muons. Then, any ``wrong sign muons''
 (negatively-charged muons) detected are an unambiguous proof of electron neutrino oscillations, in the $\nue\to\numu$ channel. This is, precisely, the golden channel, which will be shown to be essential for the goals of the present study.
We will consider a neutrino factory  providing $10^{21}$ $\mu^{+}$ and $\mu^{-}$ of $50$ GeV and a $40$ kton iron magnetized calorimeter detector. We have exploited the dependence of the signal in the neutrino energy assuming 5 energy bins~\cite{golden}. Hereafter, we will refer to this setup as the standard setup.

The appearance of ``wrong-sign muons'' at three reference baselines ($732$ Km, $2810$ Km and  $7332$ km) has been considered. Realistic background and efficiencies for the proposed $40$ kton iron magnetized calorimeter detector, see Ref~~\cite{ans}, as well as accurate matter effects along the neutrino path, have been included in our numerical analysis as well~\cite{golden}.

\section{Measurement of $\theta_{13}$ \emph{and} CP-$\delta$ at a neutrino factory through the golden channels.}
 In this section we will show that the most promising way to determine the unknown parameters $\delta$ and $\tetaot$ is through the detection of the subleading transitions $\nue \to \numu$ and $\nuebar \to \numubar$ by using the golden signature of wrong sign muons~\cite{golden}. Defining $\Delta_{ij} \equiv \frac{\Delta m^2_{ij}}{2 E}$, a convenient and
precise approximation is obtained by expanding to second order 
in the following small parameters: 
$\tetaot$, $\Delta_{12}/\Delta_{23}$, $\Delta_{12}/A$ and $\Delta_{12} \, L$. 
The result is (details of the calculation can be found in Ref~\cite{golden}):
\bea
P_{\nu_ e \nu_\mu ( \bar \nu_e \bar \nu_\mu ) } & = & 
s_{23}^2 \sin^2 2 \tetaot \, \left ( \frac{ \delot }{ \tilde B_\mp } \right )^2
   \, \sin^2 \left( \frac{ \tilde B_\mp \, L}{2} \right) \, + \, 
c_{23}^2 \sin^2 2 \theta_{12} \, \left( \frac{ \Delta_{12} }{A} \right )^2 
   \, \sin^2 \left( \frac{A \, L}{2} \right ) \nn \\
& + & \label{approxprob}
\tilde J \; \frac{ \Delta_{12} }{A} \, \frac{ \delot }{ \tilde B_\mp } 
   \, \sin \left( \frac{ A L}{2}\right) 
   \, \sin \left( \frac{\tilde B_{\mp} L}{2}\right) 
   \, \cos \left( \pm \delta - \frac{ \delot \, L}{2} \right ) \, , 
\label{eqn:hastaelmogno}
\eea
where $L$ is the baseline, $\tilde B_\mp \equiv |A \mp \delot|$ and the 
matter parameter $A$ is defined in terms of the average electron 
number density, $n_e(L)$,  as $A \equiv \sqrt{2} \, G_F \, n_e(L)$, 
where the $L$-dependence will be taken from Ref~\cite{quigg}.
The $\tilde J$ parameter is defined as 
\be
 \tilde J \equiv \cos \theta_{13} \; \sin 2 \theta_{13}\; \sin 2 \theta_{23}\;
 \sin 2 \theta_{12}.
\ee
After a careful exploration of the energy and baseline dependence of the different terms in Eq.~(\ref{eqn:hastaelmogno})~\cite{golden}, and a detailed study of the CP-asymmetry as defined in Ref~\cite{dgh,golden,pepe1,cuatro,CP8}: 
\begin{center}
\begin{equation}
{\bar A}^{CP}_{e\mu} = \frac{ \{ N [\mu^-] / N_o [e^-] \}_+ 
                            - \{ N [\mu^+] / N_o [e^+] \}_-}{
                              \{ N [\mu^-] / N_o [e^-] \}_+ 
                            + \{ N [\mu^+] / N_o [e^+] \}_- } \, , 
\end{equation}
\end{center}
and after substracting the fake CP violating effects induced by matter, it turns out that the optimal distance to be sensitive to the CP-phase $\delta$ at a future NF exploiting muons with energies $E_\mu=50$ GeV is $\mathcal{O}$ ($3000$) Km~\cite{golden}.

In the limit  $A\rightarrow 0$, 
the expression Eq.~(\ref{eqn:hastaelmogno}) reduces to the simple formulae in vacuum~\cite{golden} 
\bea
P_{\nu_ e\nu_\mu ( \bar \nu_e \bar \nu_\mu ) } & = & 
s_{23}^2 \, \sin^2 2 \tetaot \, \sin^2 \left ( \frac{\delot \, L}{2} \right ) + 
c_{23}^2 \, \sin^2 2 \theta_{12} \, \sin^2 \left( \frac{ \Delta_{12} \, L}{2} \right ) \nn \\
& + & \tilde J \, \cos \left ( \pm \delta - \frac{ \delot \, L}{2} \right ) \;
\frac{ \Delta_{12} \, L}{2} \sin \left ( \frac{ \delot \, L}{2} \right ).  
\label{eqn:vacexpand} 
\eea

As in Ref~\cite{burguet,burguet2}, we will denote the three terms 
in Eq.~(\ref{eqn:vacexpand}), atmospheric, solar and interference, by $P^{atm}$, $P^{sol}$ and $P^{inter}_{\nu ( \bar \nu) }$, respectively. When $\theta_{13}$ is 
relatively large, the probability is dominated
by the atmospheric term. We will 
refer to this situation as the atmospheric regime. Conversely, when 
$\theta_{13}$ is very small, the solar 
term dominates $P^{sol} \gg P^{atm}_{\nu (\bar \nu )}$. This is the solar
regime. The interference term is the only one which contains the CP phase $\delta$, and it is the only one which differs for neutrinos and antineutrinos.
The  vacuum approximation Eq.~(\ref{eqn:vacexpand}) should be excellent in the SB scenario with a baseline of a few hundreds of kilometers,  while in practice it also gives a good indication for the results at the short ($732$ km) and intermediate ($2000$--$3000$km) baselines of a NF. 
 
\begin{center}
\begin{figure}[h]
\centerline{\psfig{file=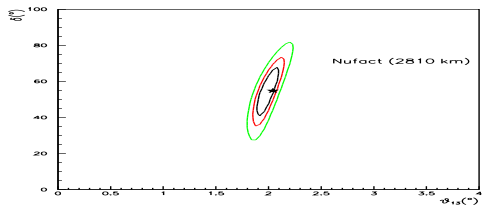, height=3cm, width=7cm}}
\vspace*{8pt}
\caption{\textit{$68.5\%$, $90\%$ and $99\%$ contours resulting from the fits at $L = 2810$ km for $\delta = 54^\circ$ and $\theta_{13}= 2^\circ$.}}
\label{fig:little}
\end{figure}
\end{center}

The next step is to  perform an exhaustive numerical treatment. All numerical results simulated have been obtained with the exact formulae for the oscillation probabilities~\cite{golden,burguet,burguet2}. We thus performed $\chi^{2}$ fit analysis to the simultaneous extraction of $\delta$ and $\theta_{13}$ at a future neutrino factory complex assuming the standard setup described in the previous section. Unless specified
 otherwise, we take the following central values for the remaining oscillation parameters: $\sin^{2} 2 \theta_{12} \cdot \Delta m^2_{12} = 1 \times 10^{-4}$ eV$^2$,  $|\Delta m^2_{23}| = 2.5 \times 10^{-3}$ eV$^2$ and  $\sin^{2} 2 \theta_{23}=1$ . The $\chi^{2}$ for a fixed baseline $\lambda$ is defined as: 
\begin{equation}
\chi_\lambda^2 = \sum_{i,j} \sum_{p,p'} \; (n^\lambda_{i,p} - N^\lambda_{i,p}) C_{i,p:,j,p'}^{-1} (n^\lambda_{j,p'} - N^\lambda_{j,p'})\,,
\end{equation}
 where $N^\lambda_{i,\pm}$ is the predicted number of wrong-sign muons for
 a certain oscillation hypothesis, $n^\lambda_{i,p}$ are the simulated ``data'' from a Gaussian or Poisson smearing and $C$ is the $2 N_{bin} \times 2 N_{bin}$ covariance matrix, that for this illustration only contains statistical errors. We can safely neglect the effects on the fit induced by the
 uncertainties in the remaining oscillation parameters. The impact of the projected uncertainties on the knowledge of the solar and atmospheric parameters and on the matter density at the time of the NF has
 been considered, and was found that these uncertainties do not significantly
 affect the global fits~\cite{burguet}. 
In Fig.~\ref{fig:little} we show the $68.5\%$, $90\%$ and $99\%$ contours for a $\chi^2$ fit to the data from a future NF with the detector of Ref ~\cite{ans} located at a baseline $L=2810$ km. The ``true'' parameter values that we have chosen for this example are depicted in the figure with a star, that is, $\delta=54^{\circ}$ and  $\tetaot=2^\circ$. 
By observing the results presented in Fig.~\ref{fig:little}, it seems that it is possible to extract simultaneously $\theta_{13}$ and $\delta$ with very good precision at a future NF with the standard setup and the far detector located at distances $\sim 3000$ Km. However, as the reader can notice,
 in this example we only have considered a limited CP-phase $\delta$ range
 ($0^\circ< \delta < 90^\circ$), while the full parameter range $-180^\circ< \delta < 180^\circ$ should be explored.
\section{Degenerate solutions.}
We can ask ourselves whether it is possible to unambiguously determine $\delta$ and $\tetaot$ by measuring the transition probabilities $\nue \to \numu$ and $\nuebar \to \numubar$ at fixed neutrino energy and at just one NF baseline. The answer is no. By exploring the full (allowed) range of the $\delta$ and $\tetaot$ parameters, that is, $-180^{\circ}<\delta<180^{\circ}$ and $0^{\circ}<\tetaot<10^{\circ}$, we find, at fixed neutrino energy and at fixed baseline, the existence of degenerate solutions ($\delta^{'}$, $\theta^{'}_{13}$), that we label \emph{intrinsic degeneracies}, which give the same oscillation probabilities than the set ($\delta$, $\tetaot$) chosen by Nature~\cite{burguet}. More explicitly, if ($\theta_{13}$, $\delta$) are the values chosen by nature, the conditions
\begin{center}
\begin{equation}
\left.\matrix{
P_{\nu_e \nu_\mu} (\theta^{'}_{13}, \delta^{'}) = P_{\nu_e \nu_\mu}
(\theta_{13}, \delta)\nonumber \cr 
P_{\bar \nu_e \bar \nu_\mu} (\theta^{'}_{13}, \delta^{'}) = P_{\bar \nu_e
\bar \nu_\mu} (\theta_{13}, \delta)}
\right \}
\label{eqn:equalburguet}
\end{equation}
\end{center}
can be generically satisfied by another set ($\theta^{'}_{13}$, $\delta^{'}$). 
It has also been pointed out that other fake solutions might appear from  unresolved degeneracies in two other oscillation parameters:
\begin{enumerate}
\item At the time of the future NF, the sign of the atmospheric mass difference $\Delta m_{23}^2$ may remain unknown, that is, we would not know if the hierarchy of the neutrino mass spectrum is normal or inverted. In this particular case, $P (\theta^{'}_{13}, \delta^{'}, -\Delta m_{23}^{2}) = P (\theta_{13}, \delta, \Delta m_{23}^2)$~\cite{sign1,sign2}.
\item Disappearance experiments only give us information on $\sin^{2} 2 \theta_{23}$: is $\theta_{23}$ in the first octant, or is it in the second one, $(\pi/2 -\theta_{23}$)? . In terms of the probabilities, $P (\theta^{'}_{13}, \delta^{'}, \frac{\pi}{2}-\tetatt) = P (\theta_{13}, \delta ,\tetatt)$~\cite{sign2,th231}.
\end{enumerate}
\begin{center}
\begin{figure}[h]
\centerline{\psfig{file=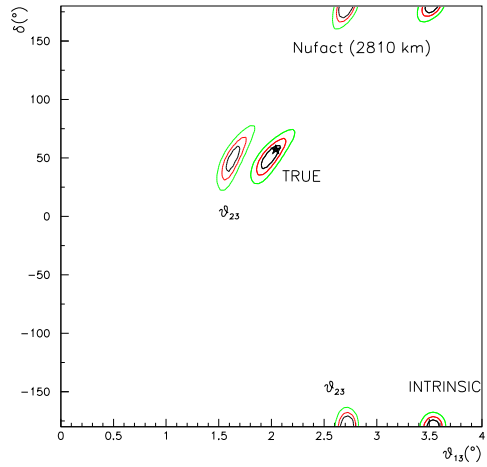, height=7cm,width=7cm}}
\vspace*{8pt}
\caption{\textit{$68.5\%$, $90\%$ and $99\%$ contours
resulting from the fits at $L = 2810$ km for $\delta = 54^\circ$ and $\theta_{13}= 2^\circ$. Three fake solutions appear, in addition to the true one. The degeneracy corresponding to the case of a global fit to data with the wrong choice of the sign of $\Delta m^2_{23}$ and of the $\tetatt$ octant is not depicted.}}
\label{fig:deg}
\end{figure}
\end{center}

All these ambiguities complicate the experimental determination of $\delta$ and
$\tetaot$. To illustrate in brief the degeneracy problem, imagine as an example that the values chosen by Nature are
$\delta=54^{\circ}$ and $\tetaot=2^{\circ}$, and consider the NF fluxes
with a detector of the type discussed in Ref~\cite{ans} located at
$L=2810$ km. A global fit of the experimental data using the spectral
information would result in a cluster of solutions some of which are 
depicted in the Fig.~\ref{fig:deg}. A constellation of fake solutions 
($\theta^{'}_{13},\delta^{'}$) surrounds the true one: those induced by the 
intrinsic degeneracy and by the $\theta_{23}$ octant ambiguity are shown, 
whereas those coming from the ambiguity in the sign of $\Delta m^2_{23}$ 
are absent for the particular case analyzed here due to the presence of 
sizable matter effects (as it is well known, if the atmospheric mass splitting is positive, i.e, normal spectrum, the oscillation probabilities of neutrinos (antineutrinos) is enhanced (depleted)). 

Finally, the combined degeneracy that would arise 
when performing a fit with both the wrong choices of the 
sign$(\Delta m^2_{23})$ and of the $\tetatt$ octant is not depicted. 
From the results shown in Fig.~\ref{fig:deg}, we see that we would not be 
able to determine whether $\delta \simeq 54^\circ$ (CP is violated) or 
$\delta \simeq 180^\circ$ (CP is conserved)!

\section{Resolving the degeneracies}
In Ref~\cite{golden} it was pointed out that some of the degeneracies listed above could be eliminated with sufficient energy or baseline spectral
 information. In practice, however, the spectral information has been
 shown to be not strong enough to resolve degeneracies with a single detector,
 once that statistical
 errors and realistic efficiencies and backgrounds are taken into account.
A lot of work has been thus devoted to resolve the degeneracies by exploiting the different neutrino energy and baseline dependence of two (or more) LBL experiments.
For instance, it has been suggested in Ref~\cite{burguet} to combine the
 results of the optimal NF baseline (${\cal O}(3000)$ km), with other NF
 baselines~\footnote{If lower energy threshold detectors are a viable option,
 the degeneracies can be resolved at a single NF baseline~\cite{huber}.}.
An independent appearance channel at future neutrino factories, the so called silver channels~\cite{andrea1,andrea2} $\nu_e \to \nu_\tau$ and $\bar{\nu}_e \to \bar{\nu}_\tau$,  can resolve the intrinsic degeneracies, provided  that $\theta_{13}> 1^\circ$. We will exploit the potential of the silver channels in the next sections.  
A higher gamma beta beam has been recently proposed and the potential of such a novel technique has been shown to be competitive with the NF~\cite{pilar}.
 For a recent theoretical study of the parameter degeneracies, and the perspectives for resolving them through the combination of data from NF, SB and/or beta beam, see Ref~\cite{stef1}. The physics potential of the combination of future beta beams and SB facilities has been carefully analyzed in Ref~\cite{stef2}.

If the value of $\theta_{13}$ turns out to be not very small, combinations of future long baseline experiments can also help enormously~\cite{barger,min1,min2,pom2}. Another possibility is to exploit the data from neutrino reactor experiments, one of the best ways to deal with the degeneracy associated with the $\theta_{23}$ ambiguity~\cite{min3,mic}.

\subsection{Exploiting the different $E$, $L$ of superbeam experiments}

Using the approximate formulae of Eq.~(\ref{eqn:vacexpand}), it is easy to 
find the expression for the {\it intrinsic} degeneracies in the atmospheric and solar regimes. In Ref~\cite{burguet} the general results including matter effects are given.  Solutions in vacuum are derived in Ref~\cite{burguet2}. These solutions are
 easier to understand and are a good approximation for the baselines
 relevant to neutrino superbeams. In the present study we review in detail the case of the intrinsic degeneracies. We then summarize the situation for the sign and $\theta_{23}$-octant degeneracies, by noticing that there exists a common pattern in these fake solutions.
  
For instance, we show that for $\theta_{13}$ sufficiently large and in the vacuum approximation, apart from  the true solution, there is a fake one at~\cite{burguet,burguet2}
\bea
\delta^{'}&\simeq&\pi -\delta,\nonumber\\
\theta^{'}_{13}&\simeq & \theta_{13} +
\cos\delta\ \sin 2\theta_{12} \;\sol \cot \theta_{23} \cot\left(\atmos\right) \; \;.
\label{eqn:atmburguet}
\eea 
Note that for the values $\delta =-90^\circ, 90^\circ$, the fake solution
 disappears. 
Typically,  $\cot\left(\atmos\right)$ has on average opposite sign for the proposed SB and NF setups\footnote{Clearly the parameters for these setups are not fixed yet and might be modified conveniently in the final designs.}, for $\Delta m^2_{23}=0.003$ eV$^2$, as can be seen in Table~\ref{cotvalues}. \\
\begin{table}[h]
\tbl{Parameters for possible superbeam, neutrino factory, and $\beta$-beam setups.}
{\begin{tabular}{@{}cccc@{}} \toprule
&$\langle E\rangle$ ({\rm GeV}) & L ({\rm km}) & $\cot\left({\Delta m^2_{23} L \over 4 E}\right)$\\ \colrule
{\rm SB-SPL}&0.25&130&{-}0.43\\
\rm{JHF-off-axis}&0.7&295&{-}0.03\\
\rm{{NF}}@732&30&732&{+}10.7\\
\rm{{NF}}@2810&30&2810&{+}2.68\\
$\beta${\rm-beam}&0.35&130&{+}0.17\\
\botrule
\end{tabular}
\label{cotvalues} 
}
\end{table}

When $\theta_{13} \rightarrow 0$ and in the vacuum approximation, the intrinsic degeneracy is 
independent of $\delta$~\cite{burguet,burguet2}:
\begin{center}
\begin{equation}
\left.\matrix{
\textrm{if}\ \cot\left(\atmos\right)  > 0\  \textrm{then} \ \delta^{'}\simeq\ \pi \cr
\textrm{if}\ \cot\left(\atmos\right)  < 0\  \textrm{then} \  \delta^{'}\simeq\ 0}
\right \} \;\;\; \theta^{'}_{13}\simeq \sin 2 \theta_{12}\  \sol \ |\cot\theta_{23}\ \cot\left(\atmos\right) \ |.
 \label{eqn:solarburguet}
\end{equation}
\end{center}

This solution is referred to in the literature as the $\tetaot =0$-mimicking
 solution, and occurs because there is a value of $\theta^{'}_{13}$ for which there 
is an exact cancellation of the atmospheric and
interference terms in both the neutrino and antineutrino probabilities 
simultaneously, with $\sin \delta'=0$~\cite{burguet,burguet2}. 

The $\theta^{'}_{13} -\tetaot$ difference of the fake solution depends strongly on the baseline and the neutrino energy through the ratio $L/E$, so the combination of the results
of two experiments with a different value for this ratio should be able 
to resolve these degeneracies. Even more important is 
that, for small $\tetaot$, $\delta'$ may differ by $180^\circ$ if the two facilities have opposite 
sign for  $\cot\left(\atmos\right)$, see Eqs.~(\ref{eqn:solarburguet}). 
 We note that $\cot\left(\atmos\right)=0$ at the maximum of atmospheric
 neutrino oscillations in vacuum.
For the NF setups, the sign of $\cot\left(\atmos\right)$ is clearly 
positive (see Tab.~\ref{cotvalues}), since the measurement of CP violation requires, because of the large matter effects, a baseline considerably shorter than that corresponding to the maximum of the
atmospheric oscillation (in vacuum). In the SB scenario on the other
hand, because of the smaller $\langle E\rangle$, matter effects are small at
the maximum of the atmospheric oscillation, which then becomes the optimal 
baseline for CP violation studies, and therefore $\cot\left(\atmos\right)$
 should be chosen close to zero.
 It is then not very difficult to ensure that $\cot\left(\atmos\right)$ be dominantly negative in this case\footnote{Note however that most neutrino beams are generally wide-band beams in energy 
so it is necessary for this argument to hold that most of the
 interactions originate from neutrinos with an energy giving the
 appropriate $\cot\left(\atmos\right)$ sign.
 The results of the fits performed in Ref~\cite{burguet2} indicate that this is the case in the two facilities (NF and SPL-SB) 
that are considered in detail in this paper. 
}, which results in an optimal complementarity of the two facilities in resolving degeneracies. 

Although the combination of the data from two SB facilities with different $E /L  \sim \Delta m^2_{23}$ could also a priori overcome the degeneracies, SB projects are in general planned to exploit data on or nearby the atmospheric
 oscillation maximum and, therefore, the differences in their $E / L$ are not large enough to fully resolve the degeneracies.

\begin{center}
\begin{figure}[h]
\centerline{\psfig{file=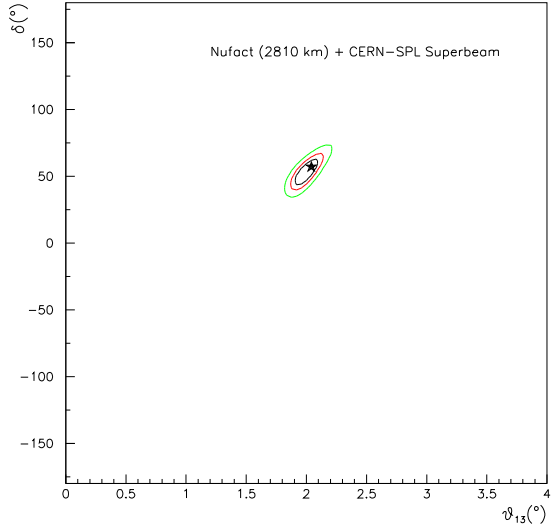, height=8cm,width=8cm}}
\vspace*{8pt}
\caption{\textit{Fits combining the results from the SPL-SB facility and from a neutrino factory 
baseline at $L= 2810$ km for $\delta = 54^\circ$ and $\theta_{13}= 2^\circ$. Notice that the fake intrinsic solutions have completely 
disappeared at the $3\sigma$ CL in the combination.}} 
\label{fig:degarr}
\end{figure}
\end{center}

 A detailed combined analysis of the results from a neutrino factory ~\cite{golden,burguet} and those from a superbeam ~\cite{sb} facility has also
 been performed. The  combination of superbeams with the optimal NF baseline of $L=2810$ km has been shown ~\cite{burguet} to be sufficient to eliminate 
 all the fake solutions even in the solar regime. For the sake of illustration, consider the values $\delta =54^{\circ}$ and  $\tetaot=2^\circ$. After combining the data obtained at a NF baseline of $L=2810$ km and the data from the SPL-SB facility, the fit clearly selects the correct solution, see Fig.~\ref{fig:degarr}.

Evidently, this example is for a rather high value of $\tetaot$, that is
 $\tetaot=2^\circ$. The parameter space down to very low values of $\tetaot$ is
 explored in detail in Ref~\cite{burguet2}, where it is shown that the fake
 solutions associated with the sign of the atmospheric mass difference and $\tetatt$ ambiguities can be grouped in two sets:  those close to nature's true values (solutions of type I) and those related to the intrinsic fake solution (solutions of type II). Generically solutions of type I show a milder $L/E$ dependence  and are thus more difficult to eliminate through the combination of the data from the two facilities considered here.  We study all the fake solutions both analytically and through
 numerical simulations, and here we summarize the expected results after considering both experiments:  
\begin{itemize}
\item The intrinsic degeneracies disappear after the NF+SB combination down
 to the sensitivity limit, for all three reference NF baselines considered
 here. If the other degeneracies are not considered, the sensitivity
 limit is $\tetaot\sim 0.3^{\circ}$ for a medium baseline NF ($L=2810$ km) and $\tetaot\sim 0.6^{\circ}$ for a short baseline NF ($L=732$ km). Although this short distance is below $\sim{\cal O}(1000)$ km, and therefore it is not sensitive to CP-Violation effects\footnote{We are considering neutrino energies of several tens of GeV.}, it is a very interesting distance after combining its results from those from the SPL-SB facility.

\item The degeneracies that arise due to the sign$(\Delta m^2_{23})$ ambiguity can be resolved by combining the results from a NF with $L=2810$ km and those from the SPL-SB facilities for $\tetaot\ge 1^\circ$. For shorter baselines ($L=732$ km) these fake solutions can be resolved after the combination for values of $\tetaot$ near its upper present bound of $\tetaot <10^{\circ}$, given by CHOOZ~\cite{chooz}. At very small values of $\tetaot$, the sign of $\Delta m^2_{23}$
 remains an ambiguity, but it does not interfere much with the determination of $\tetaot$ (in this particular case, $\theta^{'}_{13} =\tetaot$) and with the measurement of leptonic CP violation, since $\delta '=180^{\circ}-\delta$. The former implies that $\sin\delta '=\sin \delta$.

\item The degeneracies due to the $(\tetatt,\pi/2 - \tetatt)$ ambiguity are difficult to resolve and they can interfere with the measurement of $\tetaot$ and $\delta$. The combination of NF and SB experiments helps enormously in minimizing the bias in the extraction of $\tetaot$ and $\delta$, though. However, as we will explore in the next sections, the analysis of the silver channels
 can help in this respect.
\end{itemize}

\subsection{Exploiting the different channels at the neutrino factory}
In the previous section we have pointed out that the degeneracies associated with the $\theta_{23}$ octant ambiguity are not fully resolved after the combination of the data from NF and SB experiments. Particularly difficult to overcome are type I solutions~\footnote{The location of this degeneracy can lie far apart from the true values. In particular, if the true value of $\delta$ is CP-violating,
 this fake solution appears to be consistent with CP-conservation.} since these
 solutions are nearly $L/E$ independent. 
However, as first noticed in Ref~\cite{andrea1}, an additional measurement of an independent appearance channel, $\nu_e\rightarrow\nu_\tau$  and $\bar{\nu}_e\leftrightarrow\bar{\nu}_\tau$ (silver channels) greatly helps in resolving
 the fake solution of type I associated with the $\tetatt$-ambiguity, since the locations of the fake solutions that arise from the data analysis of silver and golden channels differ substantially.

Consider the approximate oscillation probabilities~\cite{golden,andrea1} in vacuum for $\nu_e\rightarrow \nu_\tau$ 
( $\bar \nu_e \rightarrow \bar \nu_ \tau$):
\bea
P_{\nu_ e\nu_\tau ( \bar \nu_e \bar \nu_\tau ) } & = & 
c_{23}^2 \, \sin^2 2 \tetaot \, \sin^2 \left(\atmos\right)  + 
s_{23}^2 \, \sin^2 2 \theta_{12} \, \left( \sol\right)^2 \nn \\
& - & \tilde J \, \cos \left ( \pm \delta - \atmos \right ) \;
 \sol \sin \atmos .
\label{eqn:vacexpandetau} 
\eea
They differ from those in Eq.~(\ref{eqn:vacexpand}) by the interchange $\theta_{23} \rightarrow \pi/2 -\theta_{23}$ and by a change in the sign of the interference term. As a result, in the atmospheric regime, the location of the fake solutions related to the $\theta_{23}$ ambiguity is opposite in sign
 for $\nu_e\rightarrow \nu_\tau$ and $\bar{\nu}_e\rightarrow \bar{\nu}_\tau$ oscillations and for 
 $\nu_e\rightarrow \nu_\mu$ and $\bar{\nu}_e \rightarrow \bar{\nu}_\mu$ oscillations. In the solar regime, on the other hand, the solution of type I for the $\nu_\tau$ appearance measurement coincides with the one for $\nu_\mu$ appearance, while solutions of type II are different, as shown in Ref~\cite{burguet2}.
For the analysis of the silver channel we assume the setup of the Opera proposal~\cite{opera}, with one fixed baseline, $732$ km, i.e. the distance from CERN to Gran Sasso Laboratories in Italy, and a 4 Kton lead plus emulsion detector with spectrometers. A dedicated analysis can be found in Ref~\cite{andrea2}.
We have found that after the combination of the results from NF golden ($L=2810$ km) and SB silver ($L=732$ km)  channels no degeneracy related to the $\tetatt$-octant ambiguity survives for $\tetaot>0.6^\circ$, at least assuming an
 ideal detector. This limit is expected to become more stringent, namely, $\tetaot>1^\circ$, when adding detector efficiencies and backgrounds. At very small values of $\tetaot$ ($\tetaot<0.6^\circ$), the $\tetatt$-octant ambiguity remains, but it does not interfere with the extraction of the two unknown parameters, $\tetaot$ and $\delta$, due to the location of the fake solution: $\tetaot^{'}\sim\tetaot$ and $\delta^{'}\sim \pi -\delta$.

\section{Getting the most from the combination of measurements at neutrino factories and at superbeam experiments}
\begin{center}
\begin{figure}[h]
\centerline{\psfig{file=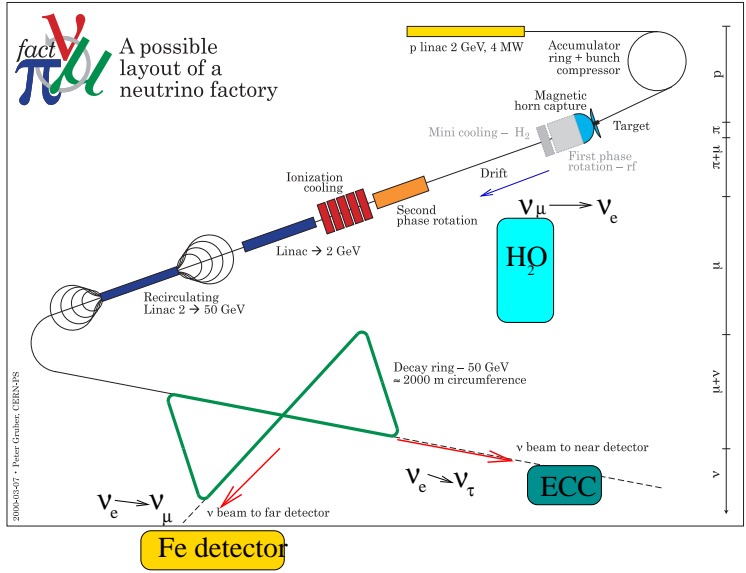,height=3in}}
\vspace*{8pt}
\caption{\textit{The complex of NF (golden and silver channels) and the SPL-SB experiments.}}
\label{fig:schedule}
\end{figure}
\end{center}

\begin{center}
\begin{figure}[h]
\centerline{\psfig{file=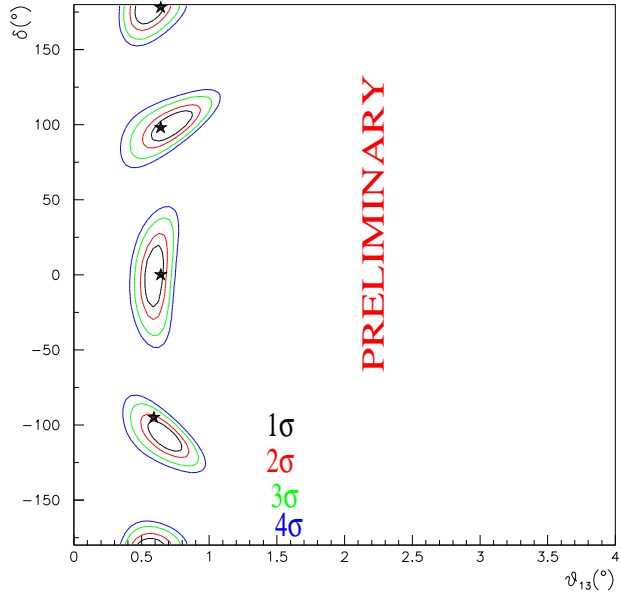,height=9cm, width=9cm}}
\vspace*{8pt}
\caption{\textit{Simultaneous extraction of $\delta$ and $\theta_{13}$ after the combination of three experiments described in the text.}} 
\label{fig:final}
\end{figure}
\end{center}

We have considered~\cite{moriond} the impact of three simultaneous experiments (see Fig.~\ref{fig:schedule}) with 2 years running in the $\pi^{+}$ polarity and 10 years running in the $\pi^{-}$ polarity~\cite{prep}.
 Simulated data from golden channels in a SB experiment, and
 from golden and silver channels in a NF experiment, have been combined.
 As the first stage of a detailed study with a realistic experimental setup~\cite{prep} we consider here an ideal situation, neglecting backgrounds and efficiencies for an emulsion cloud chamber (ECC) detector. 
In Fig.~\ref{fig:final} we show preliminary results for the simultaneous measurement of $\theta_{13}$ and $\delta$ for true values $\theta_{13}=0.6 ^\circ$ and four possible values for the CP phase $\delta= 90, 0, -90$ and $180$ degrees. 

\section{Conclusions}

It is a very interesting era for neutrino physics. New physics is revealed
 at a steady pace from a wealth of neutrino oscillation measurements.
 One of our major challenges is to search for a theoretical model to accommodate in a natural way the new physics scale that the observations
 suggest.
A measurement of the absolute mass for one neutrino would allow to guide us towards this new physics scale. Neutrino oscillation physics is already able to determine the mass differences and the mixing parameters between neutrino flavors and contribute potentially to the overall understanding of the origin of fermion masses.  There exist several ongoing neutrino oscillation experiments which exploit solar, reactor, atmospheric and accelerator neutrinos  and there are some planned experiments as well (mostly accelerator-based). All of them  have inaugurated a precision era in Neutrino Physics. Here we have focused in one
 of the major challenges ahead, that is the measurement of
 the angle $\tetaot$ and the CP-phase $\delta$. The difficulties
 associated with extracting these two parameters have led us to consider 
 the potential of Superbeam (SB) and Neutrino Factory (NF) facilities.

Short-term superbeam and longer-term neutrino factory experiments can be
 viewed as sequential steps towards the same physics goals, and not as two alternative options. In this perspective, it is natural to combine their expected
 results. We have thus shown the enormous potential of combining the data from the SB and NF facilities with realistic setups, to eliminate the degeneracies in the simultaneous measurement of $\delta$ and $\tetaot$. 

The only degeneracy that survives after the combination of SB and NF data is
 the one associated with the $\tetatt$ ambiguity. We have shown that it is possible to eliminate this degeneracy through the combination of NF golden and silver channels down to values of $\tetaot\sim 0.6^\circ$ in the present analysis,
 considering an ideal detector for the silver channel. A realistic experimental scenario is under study~\cite{prep}.

It is very important to note that there exist several alternative experimental setups that could help enormously in disentangling the neutrino puzzle: Nova~\cite{nova}, T2K~\cite{t2k}, reactor experiments and beta-beams facilities, whose potential is not discussed here. 

All the latter experimental options have to be thoroughly explored in order to ascertain the ultimate precision in the determination of the detailed pattern of neutrino mass differences and mixing angles, a prerequisite to understand their origin and their relationship to the analogous parameters in the quark sector. The NF (golden \textit{and} silver channels) plus its predecessor, the SB experiment, would provide the key to fulfill this goal.

\section*{Acknowledgements}
The author would like to thank J.~Burguet-Castell, A.~Cervera, A.~Donini, M.~B.~Gavela, J.~G\'omez C\'adenas, P.~Hern\'andez and S.~Rigolin for collaboration. It is a great pleasure to thank  M.~B.~Gavela, S.~Rigolin and  M.~Sorel for useful comments and suggestions on the manuscript. Fermilab is operated by URA under DOE contract DE-AC02-76CH03000.

\end{document}